\documentclass[%
 reprint,
superscriptaddress,
 amsmath,amssymb,
 aps,
]{revtex4-2}

\usepackage{my_macros}
\usepackage{graphicx}

\begin{document}

\title{Beyond Snap-Fit: Optimizing the Lifting Capabilities of a Partial Cylindrical Shell}

\author{
Grace K. Curtis}
\affiliation{Mathematical Institute, University of Oxford, Woodstock Rd, Oxford, OX2 6GG, United Kingdom}

\author{Ian M. Griffiths}
\affiliation{Mathematical Institute, University of Oxford, Woodstock Rd, Oxford, OX2 6GG, United Kingdom}

\author{Dominic Vella}
\affiliation{Mathematical Institute, University of Oxford, Woodstock Rd, Oxford, OX2 6GG, United Kingdom}
\email{dominic.vella@maths.ox.ac.uk}

\begin{abstract}
The cylindrical snap-fit is a ubiquitous fastening method that is both simple to manufacture and assemble, and yet secure. It consists of a partial cylindrical shell that `snaps' onto a cylindrical object.
We build on previous work to describe the mechanics of the cylindrical snap-fit as a naturally curved thin elastic shell placed atop a rigid cylinder; we investigate the shell's behaviour when subject to a point force pushing it onto or pulling it off the cylinder. We classify the possible contact regimes according to whether the shell has a nonzero lifting capacity. We term situations with lifting capacity `grip-fits' and show that this includes both the  snap-fit   and a `stick-fit' regime, which allows lifting despite not having the characteristic `snap'. Regimes without lifting capacity are also characterized for completeness.
We show that the different regimes may be characterized entirely by the shell/cylinder geometry and the coefficient of friction. We then consider different metrics for the lifting performance in the grip-fit regime. Our analysis reveals the trade-offs between assembly force, disassembly force, lifting force, and clamping force, providing design principles for secure lifting, easy detachment, and safe handling of fragile objects.

\end{abstract}



\maketitle

\section{Introduction}
Snap-fits are an easy-to-use attachment mechanism, valued for the speed, reversibility, and strength of attachment that they offer, but hidden within a wide variety of everyday items such as pen lids, buckle clips, and LEGO. Whilst there exist various forms of snap-fits, the general premise remains the same: a flexible component deforms to engage with the rigid mating part, before returning to its load-free or small-displacement state once joined \cite{Klahn2016snap}. 
Generating this engagement typically requires no complex external input, with the components coming together under simple compression.
Whether the intended use of the snap-fit is permanent installation or repeated disconnection, one of the key characteristics is that the assembly force is less than that of the disassembly force \cite{Bonenberger2016snaphandbook,Suri2000snapOptimize}.
This simplicity and strength of attachment make snap-fits a user-friendly and economical choice of connector, especially with their inherent reusability \cite{Bonenberger2016snaphandbook}. The aforementioned features have also  recently made snap-fits a contender for energy-absorption methods \cite{sano2023absorber,xu2023energy}. 

\begin{figure*}[ht]
\centering
\begin{overpic}[width=.96\textwidth,tics=10]{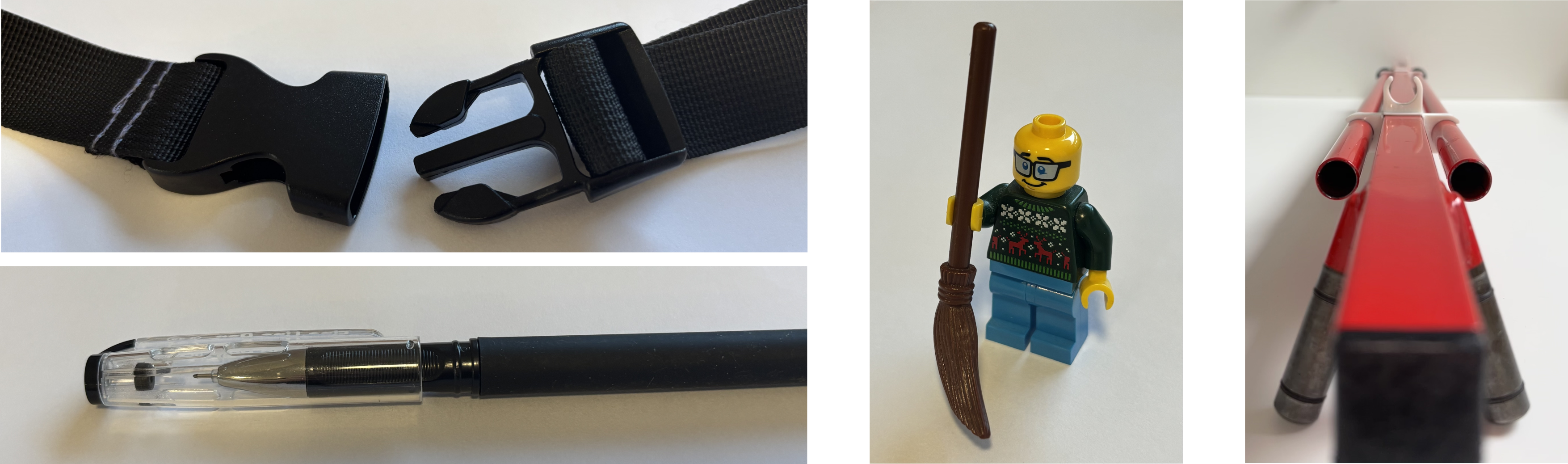}
\put(-3.5,27.5){\small{(a)}}
\put(-3.5,10.75){\small{(b)}}
\put(52,27.5){\small{(c)}}
\put(75.75,27.5){\small{(d)}}
\end{overpic}
\caption{Images showing examples of (a) cantilever snap-fit (buckle clip), (b) annular snap-fit (pen lid), and cylindrical snap-fits (the topic of this paper) for (c) a LEGO figure holding a broom and (d) to store the  legs of a collapsible bench.}
\label{fig:snap fit examples}
\end{figure*}
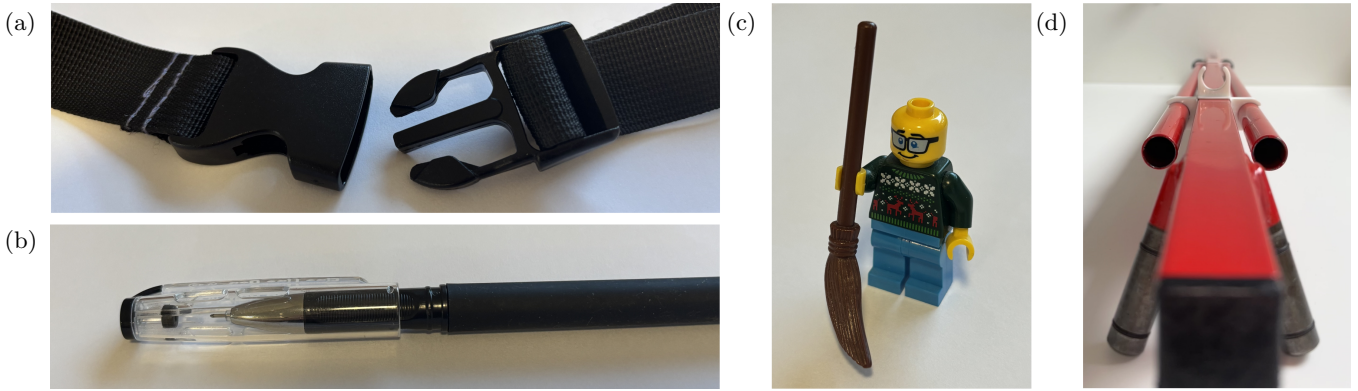

A very common and well-researched example of the snap-fit mechanism is the cantilever clip, found in many household and industrial items such as battery covers, Tupperware, buckle clips, and even automobile parts (fig.~\ref{fig:snap fit examples}(a)). Consisting of a flexible cantilever beam hook and a grooved mating component, the beam deforms until it interlocks with the groove in a (usually) stress-free state \cite{Messler1997snap}. This area has seen research into assembly and disassembly forces alongside the resulting beam stress, as well as the dependency of force on the angle of the hook-face \cite{chen2012cantilever,moses2019cantilever}.

Another, less researched snap-fit is the annular version used in aerosol lids, water bottle caps, and pen lids (fig.~\ref{fig:snap fit examples}(b)). Consisting of cylindrical parts, one flexible cap with a lip/groove and a rigid mating body with a groove/lip, the cap deforms until the lip and groove engage \cite{troughton2008handbook}. 

Perhaps one of the simplest mechanically, the cylindrical snap-fit is found in pipe clips, storage, and children's toys such as LEGO, where the mechanism is used for hands (fig.~\ref{fig:snap fit examples}(c--d)) \cite{guo2024snap,matsumoto2026sliding,yoshida2020snap}.  Consisting of a flexible partial cylindrical shell and a rigid mating cylindrical object, the shell is pushed onto the cylinder. This pushing opens the shell until it `snaps' onto the object and returns to a small-displacement state. Despite (or perhaps because of) its simplicity, the cylindrical snap-fit has been much less researched than the cantilever clip. Nevertheless, Yoshida and Wada \cite{yoshida2020snap} investigated the mechanics of the cylindrical snap-fit from the point-of-view of thin structures. They investigated the effects of shell length, radius, and the coefficient of friction on the deformation and control force, identifying two distinct domains: `snap-fit' (in which the applied force ultimately becomes negative --- the shell is pulled onto the cylinder) and `mis-fit' (in which the applied force remains positive). They also found that the edges of the shell that contact the cylinder can buckle inwards when the shell is close to a complete circle, leading to high deflections. By considering the elastica equation in the limit of small reaction forces, they produce a first-order analytic approximation to the control force, horizontal displacement and phase transitions, which they then compared to results from simulations and experiments. Guo and Sun, in \cite{guo2024snap}, offer alternative analytic results and accompanying finite-element simulations. However, neither study undertook a detailed study of how the performance of snap-fits could be optimized.

Snap-fits are often utilized to secure two components, but the advantages of the mechanism overlap with some of the desired features found in robotic grippers (specifically the speed, reusability and simplicity of attachment). As such, the cylindrical snap-fit  may represent the simplest example of a two-claw gripper \cite{samadikhoshkho2019gripper} --- provided that it operates in the industrially favourable case in which the cylinder gently slides around the cylinder without buckling. (In the nomenclature of \cite{yoshida2020snap}, this is a  `Type I' deflection, rather than the buckled (and hence larger) deflections, which they called `Type II'.)

By broadening the scope to include other forms of gripper, we are no longer limited to require that the shell `snaps' onto the object to be loaded, as required for a snap-fit. Instead, there is also the possibility of a `grip-fit', i.e.~a cylindrical shell such that, when assembled, it has a nonzero lifting capacity but without necessarily `snapping' during assembly. 
Depending on the gripper's task, one could prioritize the resulting lifting force of the shell \cite{lin2024optimizing}, or even consider the possibility of material damage by monitoring and minimizing the clamping force of such a gripper \cite{liu2018optimal}. In this article, we  expand on previous work investigating the mechanics of snap-fits to study the lifting of a rigid cylinder by a naturally curved thin cylindrical elastic shell. In particular, we focus on the broader category of grip-fit (i.e.~shells capable of lifting a load) and, for simplicity, focus on Type I (unbuckled) deformations~\cite{yoshida2020snap}. We investigate how geometric parameters affect the lifting capability under different metrics and show that the optimal gripper may actually be found beyond the snap-fit regime.

The plan of the paper is as follows: In Section 2, we introduce our model and discuss some key geometric results. In Section 3, we investigate the different regimes in our model and the criteria defining each. Our resulting grip-fit regime includes the well-known snap-fit regime, for shells that `snap' to the cylinder, but also the stick-fit regime, for shells that do not snap but still sustain a nonzero lifting capacity.  We also introduce the eject-fit regime, for shells that repel from the cylinder and have no lifting capacity. Finally, the no-fit regime describes shells that do not simply wrap around the cylinder. In Section 4, from our investigations of the different regimes, we build an understanding of the behaviour of the forces underpinning the cylindrical grip-fit, and discuss the existence of optimal parameters to suit a given lifting task. We finish by considering the clamping force of the gripper and ask whether, and when, a grip-fit can safely lift a fragile object. 

\section{Model setup}

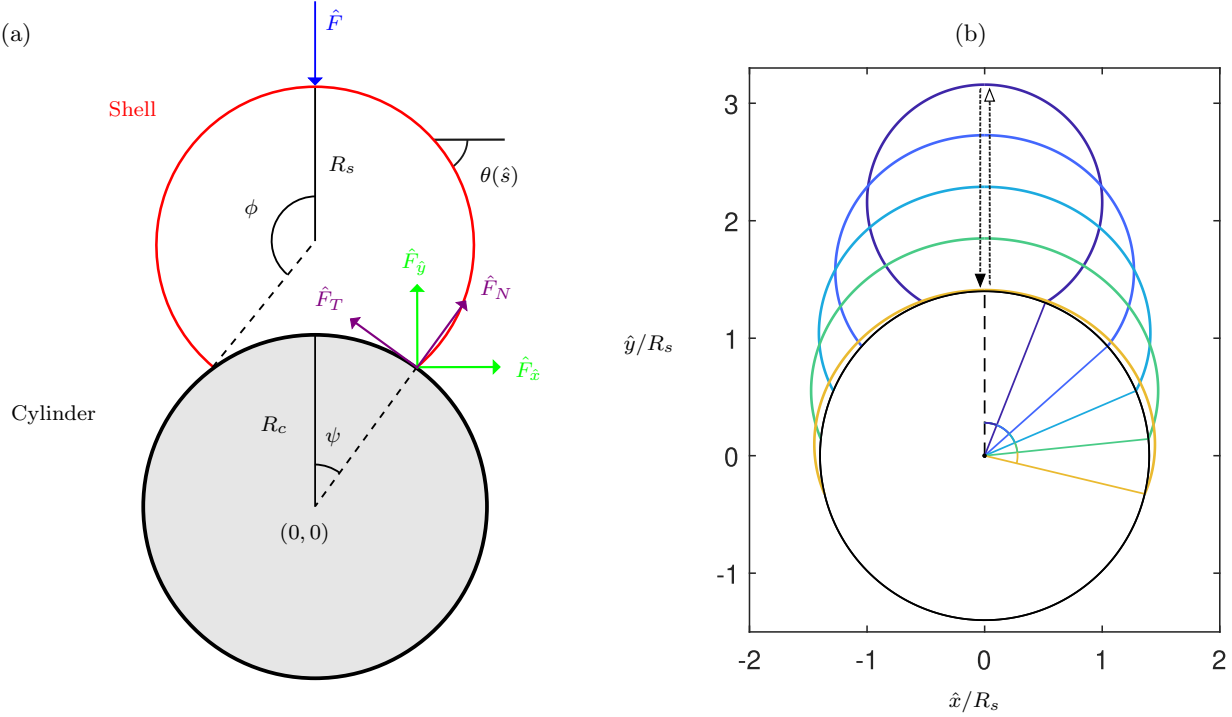
\begin{figure*}[ht]
\centering
\vspace{4mm}
\begin{overpic}[width=.8\textwidth,tics=10]{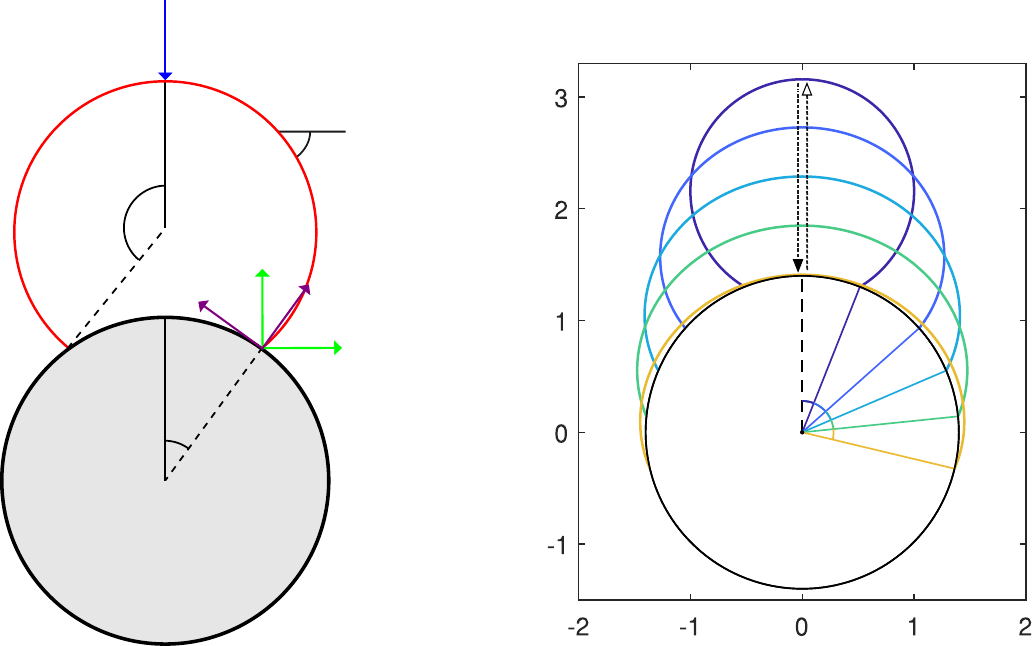}
\put(-13,59){\small{(a)}}
\put(75,59){\small{(b)}}
\put(12.75,13){\footnotesize$(0,0)$}
\put(9.5,43){\footnotesize$\phi$}
\put(17,22){\footnotesize$\psi$}
\put(17.25,47){\footnotesize$R_s$}
\put(11,23){\footnotesize$R_c$}
\put(31.25,46){\footnotesize\textcolor{black}{$\theta(\hat{s})$}}
\put(17,60){\footnotesize\textcolor{blue}{$\hat{F}$}}
\put(24,38){\footnotesize\textcolor{green}{$\hat{F}_{\hat{y}}$}}
\put(34.5,28){\footnotesize\textcolor{green}{$\hat{F}_{\hat{x}}$}}
\put(31.25,35.5){\footnotesize\textcolor{violet}{$\hat{F}_N$}}
\put(16,34.5){\footnotesize\textcolor{violet}{$\hat{F}_T$}}
\put(-12,24){\footnotesize{Cylinder}}
\put(-3,52){\footnotesize\textcolor{red}{Shell}}
\put(44.5,30.5){\footnotesize{$\hat{y}/R_s$}}
\put(74.5,-2.5){\footnotesize{$\hat{x}/R_s$}}
\end{overpic}
\vspace{2mm}
\caption{(a) Sketch of the setup for an undeformed, naturally curved elastic shell of arc length $2R_s\phi$ and radius of curvature $R_s$ atop a rigid cylinder of radius $R_c > R_s$. A localized force $\hat{F}$ is applied at the midpoint (as shown) forcing the shell to conform to the cylinder. (b)~Plots showing the shape of the shell at different points in the assembly/disassembly process (for $\alpha=R_c/R_s=1.4$, $\phi=2.6$, and $\mu=0.1$). The direction of assembly and disassembly are given by the solid and hollow arrows, respectively. }
\label{fig:snap fit sketch}
\end{figure*}

We consider the thin shell to be a naturally curved elastic beam of arc length $L=2R_s \phi$ and natural curvature $\kappa_0=1/R_s$, where $R_s$ is the radius of the shell's centre-line and $2\phi$ is the opening angle (fig.~\ref{fig:snap fit sketch}a). The endpoints of the beam are in contact with a rigid cylinder of radius $R_c > R_s$; the two contact points subtend an angle  $2\psi$ from the cylinder's centre. We assume that contact occurs only at the two endpoints of the shell because of the different curvatures of the shell and cylinder. The displacement and resulting deformation of the shell onto the cylinder occurs under the action of a localized force $\hat{F}$ acting at the midpoint of the shell.   We assume that the shell is placed symmetrically onto the cylinder, and maintains symmetry throughout the deformation. Additionally, we assume that the system is translationally invariant along the axis of the cylinder and consider quantities per unit width; for example, the line force $\hat{F}$ has units $\mathrm{N/m}$.

The coordinates $(\hat{x},\hat{y})$ are defined such that the origin lies at the centre of the cylinder. We define $\theta(s)$ as the angle the tangent to the shell makes with the positive $\hat{x}$-axis as a function of the arc length $\hat{s}$; the origin $\hat{s}=0$ is defined to be at the midpoint of the shell so that $\hat{s}=\pm R_s \phi$ at the endpoints. We note that the  angular position of the end-points, $\psi$, may be determined via the geometrical relationship
\begin{align}
    \psi=\arccos\left(\frac{\hat{y}(R_s\phi)}{R_c}\right).
\end{align}

Each contact point with the cylinder provides a reaction force with horizontal and vertical components, $\hat{F}_{\hat{x}}$ and $\hat{F}_{\hat{y}}$, respectively. Alternatively, the components of the reaction force in the direction of the normal and tangent to the surface of the cylinder are denoted $\hat{F}_N$ and $\hat{F}_T$, respectively. We will impose Coulomb friction at the contact points assuming that the shell is at the onset of sliding, i.e.
\begin{align}
\label{eqn:mu definition}
    \hat{F}_T=\mu \hat{F}_N,
\end{align}
where $\mu$ is the coefficient of friction. 

We will consider both the the shell's attachment onto (and then detachment from) the cylinder; we refer to this as assembly (and disassembly). Assembly begins with the shell in its undeformed state, placed atop the cylinder (as shown in fig.~\ref{fig:snap fit sketch}(a)). The shell is then pushed down onto the cylinder by a localized force $\hat{F}$; assembly continues until the top of the shell reaches the top of the cylinder (fig.~\ref{fig:snap fit sketch}(b)). Disassembly reverses the assembly problem: the shell is pulled from the cylinder with a change in the sign of both $\hat{F}$ and the friction coefficient. There is a resulting jump in force for nonzero $\mu$ --- it is this symmetry breaking that makes the snap-fit a useful connection method. 

\subsection{Governing equations}
We consider the same setup as \cite{guo2024snap} and \cite{yoshida2020snap}, but rather than use a discrete model for simulation or FEM, we consider a continuum elastica formulation (see, for example, \cite{Curtis2026bridging}).  Geometry relates the Cartesian coordinates $(\hat{x},\hat{y})$ of a point on the shell's centreline to the intrinsic coordinates $(\hat{s},\theta)$ via 
\abeqn{eqn:SinCosDim}{\dv{\hat{x}}{\hat{s}} = \cos\theta, \hspace{30mm} \dv{\hat{y}}{\hat{s}} = -\sin\theta.}
We exploit the left--right symmetry of the problem to consider only the half of the shell with $0 \leq \hat{s} \leq R_s\phi$. A vertical and horizontal force balance then gives
\abeqn{eqn:ForceBalanceDim}{
    \dv{\hat{s}}(\hat{n}\cos\theta-\hat{t} \sin\theta) = 0, \hspace{12mm} 
    \dv{\hat{s}}(\hat{t} \cos\theta + \hat{n}\sin\theta) = 0, 
}
respectively. Here $\hat{t}$ and $\hat{n}$ are the tension and shear forces, defined to be tangent and normal to the shell, respectively.
Finally, a moment balance gives
\begin{equation}
    \dv{\hat{m}}{\hat{s}} + \hat{n} =0,\label{eqn:MomentBalanceDim}
\end{equation}
for the bending moment $\hat{m}$. To account for the natural curvature of the shell, we impose the constitutive relation for $\hat{m}$,
\begin{equation}
     \hat{m}=-EI \left(\dv{\theta}{\hat{s}}-\frac{1}{R_s}\right),
     \label{eqn:ConstitutiveRelation}
 \end{equation} where $E$ is the Young's modulus of the shell, $I$ its second moment of inertia, and $\kappa_0=1/R_s$ its natural curvature.

To solve the system of equations \eqref{eqn:SinCosDim}--\eqref{eqn:ConstitutiveRelation}, we require one of either the force or height of the shell to be our control parameter. Since the maximum and minimum height of the shell midpoint is known in advance from geometry, we will treat this is as our control parameter. We therefore have a sixth-order system of equations with one unknown, $\hat{F}$, requiring seven boundary conditions. 
From symmetry, we require that
\abeqn{BC:symm}{\theta(0)=0, \hspace{40mm} \hat{x}(0) = 0.}
We must also have by geometry that 
\begin{align}
\label{BC:y_0}
    \hat{y}(0)=\hat{y}_0,
\end{align}
where $\hat{y}_0$ is the distance from the top of the shell to the centre of the cylinder. (We treat $\hat{y}_0$ as our control parameter, where in the undeformed initial state $\hat{y}_0=\hat{y}_0^{\mathrm{in}}$ will be determined by geometry.)
At the contact point, we must have that the bending moment vanishes --- the edge is simply supported --- and so, using \eqref{eqn:ConstitutiveRelation}, we have 
\begin{align}
\label{BC:theta}
    \left.\dv{\theta}{\hat{s}}\right|_{\hat{s}=R_s \phi}=\frac{1}{R_s}.
\end{align}
By geometry, the ends of the shell lie on the surface of the cylinder, i.e. 
\begin{align}
\label{BC:x^2+y^2}
    \hat{x}(R_s\phi)^2 + \hat{y}(R_s\phi)^2= R_c^2.
\end{align}

To impose the friction condition at the contact point in terms of $\hat{t}$, $\hat{n}$, and $\theta$, we must consider the balance of forces. A global vertical force balance gives 
\begin{align}
\label{eqn:F_y}
\hat{F}_y=\frac{\hat{F}}{2}.
\end{align}
Considering a horizontal force balance, and using \eqref{eqn:mu definition},we find that 
\begin{align}
\label{eqn:F_x}
    \hat{F}_{\hat{x}}=\frac{\hat{F}}{2} \frac{\tan\psi - \mu}{1+\mu\tan\psi}=\frac{\hat{F}}{2} \tan(\psi-\psi_\mu),
\end{align}
where $\psi_\mu=\arctan\mu$ is the \emph{friction angle}. Equations \eqref{eqn:F_y} and \eqref{eqn:F_x} may be expressed  as boundary conditions on $\hat{t}$ and $\hat{n}$ at the contact point using \eqref{eqn:mu definition} to give 
\begin{subequations}
\begin{align}
    \hat{n}(R_s\phi)=\frac{\hat{F}}{2} \frac{\cos(\psi-\theta_c)+\mu \sin(\psi - \theta_c)}{\cos(\psi)+\mu \sin(\psi)}, \\
    \hat{t}(R_s\phi)=\frac{\hat{F}}{2} \frac{\sin(\psi-\theta_c)-\mu \cos(\psi - \theta_c)}{\cos(\psi)+\mu \sin(\psi)},
\end{align} \label{BC:n and t}
\end{subequations}
where we have let $\theta_c=\theta(R_s \phi)$.

\subsection{Non-dimensionalization}
Taking the shell radius $R_s$ to be the natural length scale, we non-dimensionalize by letting
\begin{widetext}\begin{align}
    \hat{s}&=R_s s, & \hat{x}&=R_s x, & \hat{y}&=R_s y, &       
(\hat{F},\hat{n},\hat{t},\hat{F}_{\hat{x}}, \hat{F}_{\hat{y}},\hat{F}_N,\hat{F}_T) &= \frac{EI}{R_s^2} (F,n,t,F_x,F_y,F_N,F_T). \label{eqn:NonDimScaling}
\end{align}
\end{widetext}
This non-dimensionalization introduces an important dimensionless parameter: the ratio of radii $\alpha=R_c/R_s$, where we note that $\alpha>1$ for this problem. 

The dimensionless counterparts to equations \eqref{eqn:SinCosDim}--\eqref{eqn:MomentBalanceDim}, to be solved in \mbox{$0\leq s\leq\phi$}, are  
\begin{subequations}
\begin{align}
    \dv{x}{s} &= \cos\theta, \\
    \dv{y}{s} &= -\sin\theta, \\
    \dv{s}(n\cos\theta-t \sin\theta) &= 0, \\
    \dv{s}(t \cos\theta + n\sin\theta) &= 0,\\
    \dv[2]{\theta}{s} - n &= 0, 
\end{align} \label{eqn:DimensionlessSystem}%
\end{subequations}
with boundary conditions \eqref{BC:symm}--\eqref{BC:x^2+y^2} and \eqref{BC:n and t} becoming 
\abeqn{BCs:Dimensionless}{\theta(0)=0, \hspace{11mm} x(0) = 0,}
\vspace{-5mm}
\xyeqn{}{y(0) =y_0, \hspace{10mm} \dv{\theta}{s}{(1)}=1, \eqno{(\theequation{\mathrm{c,d}})}}
\vspace{-3mm}
\xyeqn{}{x^2(\phi) + y^2(\phi)= \alpha^2,\eqno{(\theequation{\mathrm{i}})}}
\vspace{-3mm}
\xyeqn{}{n(\phi)=\frac{F}{2} \frac{\cos(\psi-\theta_c)+\mu \sin(\psi - \theta_c)}{\cos(\psi)+\mu \sin(\psi)},
\eqno{(\theequation{\mathrm{j}})}}
\vspace{-1mm}
\xyeqn{}{t(\phi)=\frac{F}{2} \frac{\sin(\psi-\theta_c)-\mu \cos(\psi - \theta_c)}{\cos(\psi)+\mu \sin(\psi)},\eqno{(\theequation{\mathrm{k}})}}
where, for notational convenience, we have introduced $\theta_c:=\theta(\phi)$. 

\subsection{Numerical approach}

To solve the system numerically, we use the solver \texttt{bvp4c} in \textit{MATLAB} and we decrease $y_0$ via continuation, and find the corresponding value of $F$.

The starting point of this continuation algorithm comes from the initial point of contact, which can be determined from geometry. In particular, the angular position of the contact point is simply
\begin{equation}
    \psi=\arcsin\left(\frac{\sin\phi}{\alpha}\right),
\end{equation} while the initial value of $y_0$ is
\begin{align}
    y^{\mathrm{in}}_0=1-\cos\phi+\alpha\cos\psi.
\end{align}
Our continuation commences with $y_0=y_0^{\mathrm{in}}$ and proceeds towards the smallest possible  $y_0 \searrow \alpha$. We then reverse the displacement over increasing values of $y_0$ for the disassembly.

\section{Three fit regimes}

We find that there exist three possible fit regimes (see fig.~\ref{fig:shell_examples_and_force}(a)). We define each of these as
\begin{enumerate}
    \item Eject-fit: where, if the force is removed at any point in assembly, the shell will repel off the cylinder. 
    \item Stick-fit: where, past a critical point in assembly, if the force is removed the shell will remain stationary but never be pulled onto the cylinder. 
    \item Snap-fit: where, past a critical point in assembly, if the force is removed the shell will be pulled onto the cylinder. 
\end{enumerate}
Note we only consider cases in which there is contact at the endpoints, and do not allow for the endpoints of the shell to roll inward or slip from the surface of the cylinder. The latter cases are considered in \cite{yoshida2020snap}. 

We determine the fit regime by running the assembly process, and looking at the behaviour of the shell when $y_0 = \alpha^+$ at the start of the disassembly process. The behaviour of the shell is determined entirely by the initial geometry of the problem, characterized by $\alpha$ and $\phi$, and the coefficient of friction, $\mu$. We note that stick-fit cannot exist in the frictionless case ($\mu = 0$) as the shell will always slide.
For large enough $\phi$, we reach the regime where there does not exist a solution during displacement for which we maintain point contact at the endpoint of the shell; we denote this region `no-fit'. The corresponding value of $\phi=\phi_{\text{no-fit}}$ on the no-fit boundary also depends on the value of $\alpha$ and $\mu$. 

\subsection{Regime categorization by force}

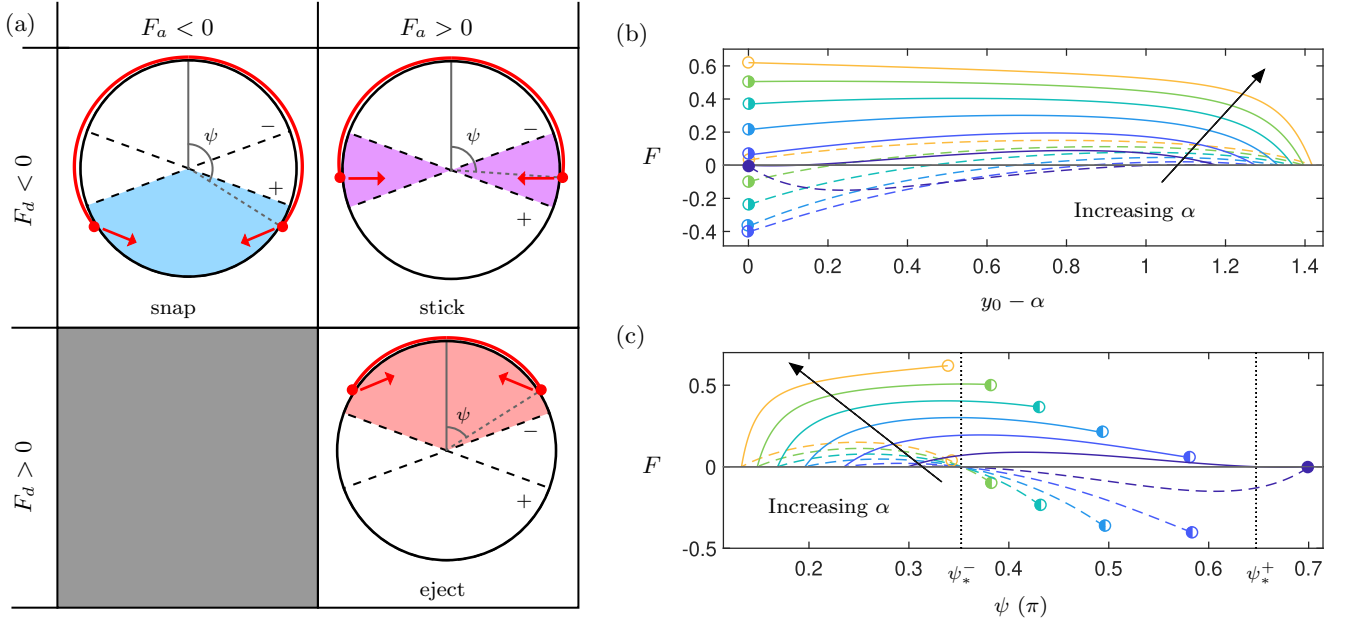
\begin{figure*}[]
\centering
\begin{overpic}[width=.98\textwidth,tics=10]{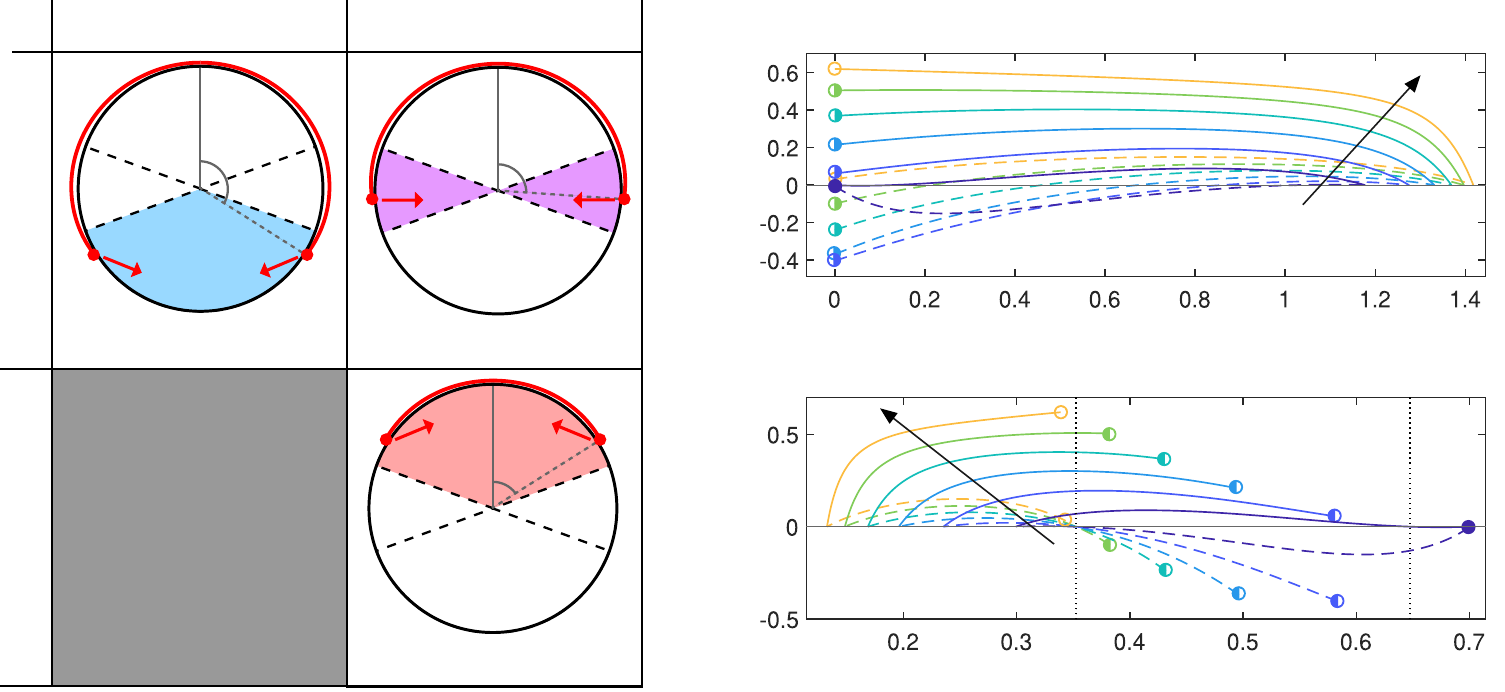}
\put(74,0){\footnotesize{$\psi$ ($\pi$)}}
\put(47.5,33.75){\small{$F$}}
\put(73,23){\footnotesize{$y_0 - \alpha$}}
\put(47.5,10.5){\small{$F$}}
\put(-0.5,43.75){\small{(a)}}
\put(45.5,43){\small{(b)}}
\put(45.5,20.25){\small{(c)}}
\put(57,7.5){\footnotesize{Increasing $\alpha$}}
\put(80,29.75){\footnotesize{Increasing $\alpha$}}
\put(93,2.5){\scriptsize{$\psi^+_*$}}
\put(70.5,2.5){\scriptsize{$\psi^-_*$}}
\put(14.5,35.5){\scriptsize{$\psi$}}
\put(19.3,31.6){\scriptsize{$+$}}
\put(18.7,36.25){\scriptsize{$-$}}
\put(34.25,35.5){\scriptsize{$\psi$}}
\put(38,29.25){\scriptsize{$+$}}
\put(38.5,36){\scriptsize{$-$}}
\put(33.5,14.5){\scriptsize{$\psi$}}
\put(38,8){\scriptsize{$+$}}
\put(38.5,13.25){\scriptsize{$-$}}
\put(10.5,22.5){\footnotesize{snap}}
\put(30.75,22.5){\footnotesize{stick}}
\put(30.75,1.25){\footnotesize{eject}}
\put(0.25,7.5){\rotatebox{90}{$F_d>0$}}
\put(0.25,29){\rotatebox{90}{$F_d<0$}}
\put(9.75,43.5){$F_a<0$}
\put(29.25,43.5){$F_a>0$}
\end{overpic}
\vspace{3.5mm}
\caption{(a) Table displaying the three different fit regimes as determined by the sign of $F_d$ and $F_a$: snap-fit, stick-fit, and eject-fit. The respective range of final contact angles $\psi$ are given by the coloured regions, with the boundaries $\psi^\pm_*$ given by the dashed lines labelled by $\pm$, respectively. The direction of movement of the shell is indicated by the reaction force arrows. The grey box highlights the unattainable case where $F_d>F_a$.
Plots of the force, $F$, against (b) the distance between the centre of the shell and the cylinder, $y_0 - \alpha$, and (c) the contact angle, $\psi$, for the assembly (solid) and disassembly (dashed) processes, for $\alpha =\{1.001,1.2,1.4,1.6,1.8,2\}$, for $\phi=2.2$ and $\mu=0.5$. The resulting fits; snap-fit, stick-fit, and eject-fit, are denoted by filled, half-filled, and empty circle markers, respectively. In (b), the critical angles $\psi^+_*$ and $\psi^-_*$ are marked on the $x$-axis by the dotted lines. }\label{fig:shell_examples_and_force}
\end{figure*}

Mathematically, we can describe each fit regime by considering the sign of the force $F$, denoted by $F_a$ and $F_d$ for assembly and disassembly, respectively. When transitioning from assembly to disassembly, the reversal in coefficient of friction $\mu \rightarrow -\mu$ leads to a jump in $F$ at $y_0 =\alpha^+$, where $F_a\geq F_d$, with equality attained when $\mu =0$. If $F_a>0$ throughout the assembly process then we are pushing the shell to the cylinder throughout; however,  if $F_a<0$ at some point, then the shell is being pulled to the cylinder.
The resulting `snapping' of the shell onto the cylinder is called snap-fit following \cite{yoshida2020snap} (see fig.~\ref{fig:shell_examples_and_force}(a,b)). 
Therefore, the sign of $F_a$ at the end of assembly ($y_0 = \alpha^+$) reveals whether we have snap-fit or not. 
Similarly, if $F_d<0$ during disassembly then the shell is being pulled to the cylinder and if $F_d>0$ then the shell is being repelled.

We can thus define the three fit regimes in the following way: 
\begin{enumerate}
    \item Eject-fit: $F_a>0$ at the end of assembly and  $F_d>0$ at the beginning of disassembly. 
    \item Stick-fit: $F_a>0$ at the end of assembly and $F_d<0$ at the beginning of disassembly.
    \item Snap-fit: $F_a<0$ at the end of assembly and $F_d<0$ at the beginning of disassembly. 
\end{enumerate}

The stick-fit regime is named because there is a negative disassembly force at the start of disassembly even though the assembly force remains positive for the entire assembly; the shell is neither snapping nor repelling. 

In the stick-fit and snap-fit cases, the shell has a nonzero lifting capacity (i.e.~a certain amount of force is still required to pull the shell from the cylinder); we have a form of grip-fit. Depending on the parameters, the magnitude of the lifting force in the stick-fit case can be larger than that in the snap-fit case (fig.~\ref{fig:shell_examples_and_force}(b)).

\subsection{Regime categorization by critical-angle}

By considering the contact angle $\psi$ at the end of assembly, we can determine the regime of the fit without having to consider the disassembly process. The sign of $F$ is crucial in determining these critical fit regime transition points. In particular, determining the value of the contact angle $\psi$ when $F=0$ allows us to determine the sign of $F$ at any given $\psi$. 
Equivalently, $F_y=0$, and so, considering $F_y/F_x=0$, we find that when $F=0$, $\psi=\psi_*^\pm$, where 
\begin{align}
    \psi_*^\pm=\frac{\pi}{2}\pm\arctan\mu.
\end{align}
We note that, for $\mu$ nonzero, there will be two values of $\psi_*$, denoted by $\psi_*^+$ for assembly (pushing the shell: $\mu>0$) and $\psi_*^-$ disassembly (pulling the shell: $\mu<0$). Note that we always have $\psi_*^+>\psi_*^-$. In other words, we have $F_a\lessgtr0$ if $\psi\gtrless\psi_*^+$, and similarly, $F_d\lessgtr0$ if $\psi\gtrless \psi_*^-$ (fig.~\ref{fig:shell_examples_and_force}(b,c)). The fit regime is defined depending on the value of $\psi$ at the point $y(0) =\alpha^+$, denoted $\psiend$, as follows:
\begin{enumerate}
\item 
\text{Eject-fit:}  \hspace{5mm}   $\psiend<\psi_*^-<\psi_*^+$.
\item
\text{Stick-fit:} \hspace{5mm}  $\psi_*^-<\psiend<\psi_*^+$. 
\item 
\text{Snap-fit:}  \hspace{5mm} $\psi_*^-<\psi_*^+<\psiend$.
\end{enumerate} 
Note that while the values of $\psi_\ast^\pm$ can be calculated immediately, determining the value of $\psi_{\rm end}$ for given material parameters requires the  solution of the full problem\footnote{Note, however, that, as suggested in \cite{yoshida2020snap}, one can approximate $\psi_{\rm end} \approx \phi/\alpha$ (i.e.~assume the shell perfectly wraps the cylinder), to gain some intuition.} because the deflection of the shell will affect the value of $\psi_{\rm end}$. 
With this angular classification of the regimes, we can now determine the fit of the shell by performing numerical simulations of the assembly process only: the value of the contact angle $\psi$ at the end of assembly, $\psiend$, compared with $\psi^+$ and $\psi^-$ determines the sign of $F_a$ and $F_d$, respectively, at $y_0 = \alpha^+$, and therefore holds the relevant information.

\subsection{Characterization of parameter space}

We will now use our model to describe the performance of a cylindrical clip as a gripper (focusing on  the lifting of cylindrical objects). If  the main purpose is to lift an object, then we are most interested in the regimes for which the shell has a nonzero lifting capacity, i.e.~the grip-fit regime. 

\begin{figure}[]
\centering
\begin{overpic}[width=.45\textwidth,tics=10]{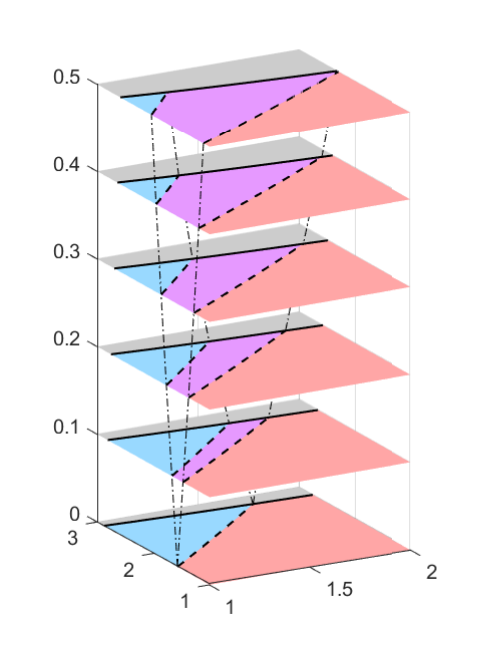}
\put(2,53.5){$\mu$}
\put(12,10){$\phi$}
\put(55,5){$\alpha$}
\put(45,102){\linethickness{0.5pt}\vector(0,-1){12}}
\put(40,103.25){\small{no-fit}}
\put(55,95){\linethickness{0.5pt}\vector(0,-1){10}}
\put(48.5,96.25){\small{eject-fit}}
\put(33,93){\linethickness{0.5pt}\vector(0,-1){10}}
\put(27,94.25){\small{stick-fit}}
\put(22,98){\linethickness{0.5pt}\vector(0,-1){14}}
\put(15.5,99.5){\small{snap-fit}}
\end{overpic}
\caption{Regime diagram showing the three possible fits for opening angles $\phi$ against ratio of radii $\alpha$, for various values of coefficient of friction $\mu$. Here, the three fit regimes are labelled by colour, and the grey region represents the no-fit regime, as labelled from above. The fit transitions are marked by the dashed lines, with the no-fit boundary denoted by a solid line. The dot-dashed lines highlight the changes in the snap-fit and stick-fit boundaries with $\mu$.}\label{fig:snap_regime_diagram}
\end{figure}
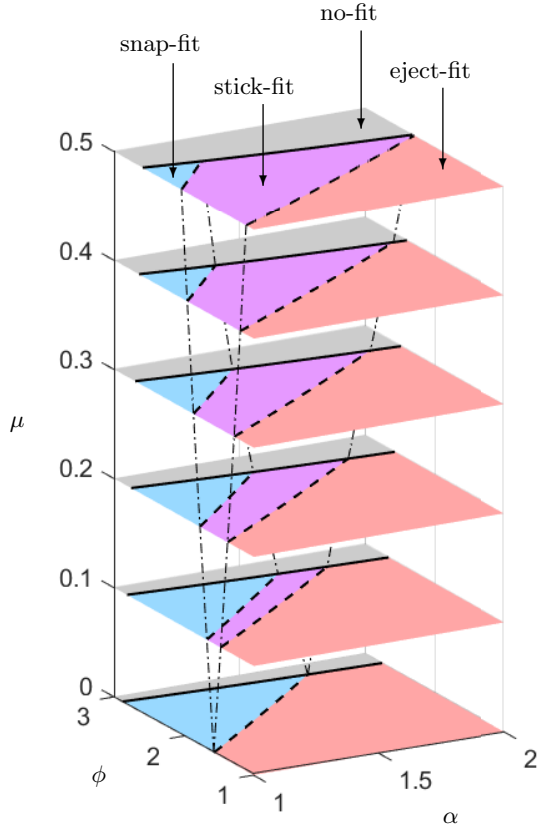

A first step is to determine which combinations of  the three control parameters ($\alpha$, $\phi$, and $\mu$) produce the different kinds of grip-fit --- snap-fit and stick-fit. Figure~\ref{fig:snap_regime_diagram} shows a regime diagram for where each fit regime is observed. (Since this parameter space is three-dimensional, we only take slices at particular values of $\mu$ to facilitate presentation; the regions vary smoothly, and so the behaviour for intermediate values of $\mu$ can easily be inferred.)

In agreement with \cite{yoshida2020snap}, we see that the size of the no-fit region increases with friction, i.e.~the boundary values of $\phi_{\text{no-fit}}(\alpha; \mu)$ decrease, as $\mu$ increases. This observation should not be surprising:  a greater frictional force makes it  more difficult for a shell with large opening angle $\phi$ to slide around the cylinder, making it more likely for the edges of the shell to roll inwards instead.
One can also see from fig.~\ref{fig:snap_regime_diagram} that as $\alpha$ increases, the range of $\phi$ values for which a snap-fit is possible decreases to zero. This indicates that, to increase the range of shells for which snap-fit is observed, one should decrease $\alpha$. 
This behaviour remains consistent between values of $\mu$, supporting the argument that the problem is highly geometrical (\cite{yoshida2020snap}). However, while the eject-fit and snap-fit regions decrease in size, the stick-fit region increases as a result. This means that, while we are less likely to see a snap-fit or eject-fit for higher values of $\mu$, the range of values of $\alpha$ and $\phi$ for which the shell has lifting capabilities has increased overall. This indicates that a larger coefficient of friction gives more freedom in choosing a shell that will be capable of lifting the desired cylinder.

\section{Load-bearing capacity and assembly forces: Different optimization criteria}

Whilst the regime diagrams provide insight into the geometric properties that control what fit regime we may see, one key function of grip-fits (particularly snap-fits) in applications is their load-bearing capacity. We define the lifting capacity of a shell, $F_d^\mathrm{max}$, to be  the largest force required to remove the shell from the cylinder, i.e.~the largest (in magnitude) negative force reached in the disassembly process. Thus, 
{\begin{align}
    F_d^\mathrm{max}= \max_{F_d<0}|F_d|. 
\end{align}
To find the lifting capacity for a given value of $\alpha$ and $\phi$, we first note that the value of the angle of contact, $\psi$, for which $F_d^\mathrm{max}$ is realised changes with $\alpha$ (fig.~\ref{fig:shell_examples_and_force}(b)). In particular, we see for most values of $\alpha$ that the largest negative disassembly force occurs at the start of disassembly, that is, at the largest value of $\psi$. However, for $1 \lesssim \alpha \lesssim 1.18$, and for sufficiently large values of $\phi$, we see that this critical $\psi$ is no longer at the start of disassembly; instead, $F_d^\mathrm{max}$ is realised at a smaller value of $\psi$. We must therefore take more care than just extracting the first disassembly force value as the most negative one. 

In fig. \ref{fig:FD_colourplots}(a) we show the raw disassembly force as $\phi$ and $\alpha$ vary for $\mu=0.5$. From this plot it appears that the maximum disassembly force is attained on the boundary between no-fit and grip-fit:  the optimal result is simply to take a shell with opening angle $\phi$ as close to the no-fit boundary, $\phi_{\text{no-fit}}$,  as possible, for a given $\alpha$. This result is intuitive in the sense that an almost complete shell is harder to remove due to the need for greater horizontal deformation to be pulled over the widest part of the cylinder. However, such a shape would also require a large assembly force, which may not be efficient for a gripper that requires repeated assembly and disassembly. We shall reconsider what is the `right' target function to optimize shortly, but note that this maximum-force behaviour is consistent across changes in the coefficient of friction,  with the magnitude of the disassembly force also increasing with $\mu$.

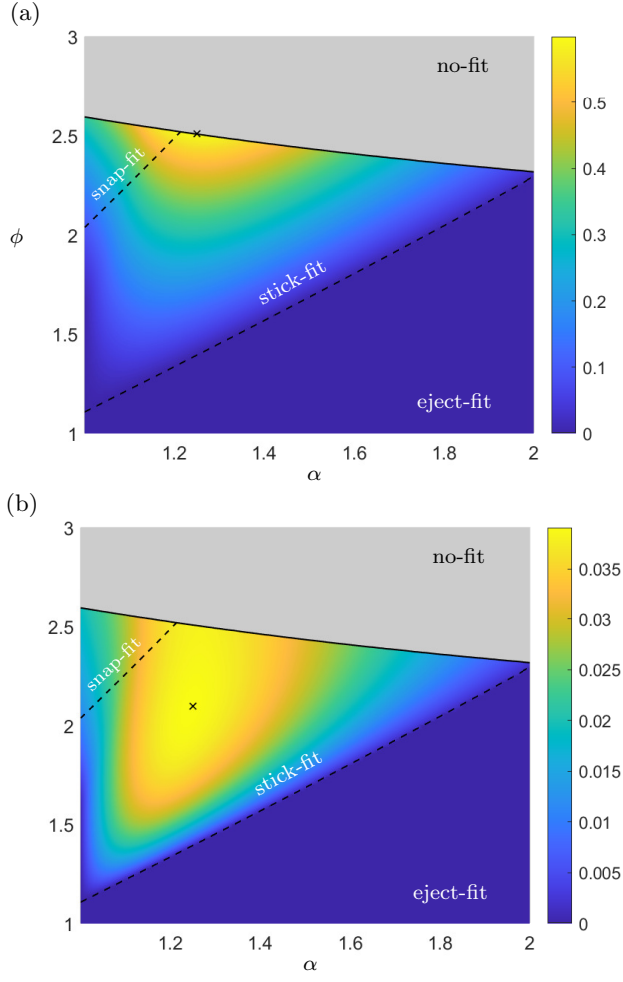
\begin{figure}[]
\begin{subfigure}{0.9\columnwidth}
\begin{overpic}[width=1.1\linewidth,tics=10]{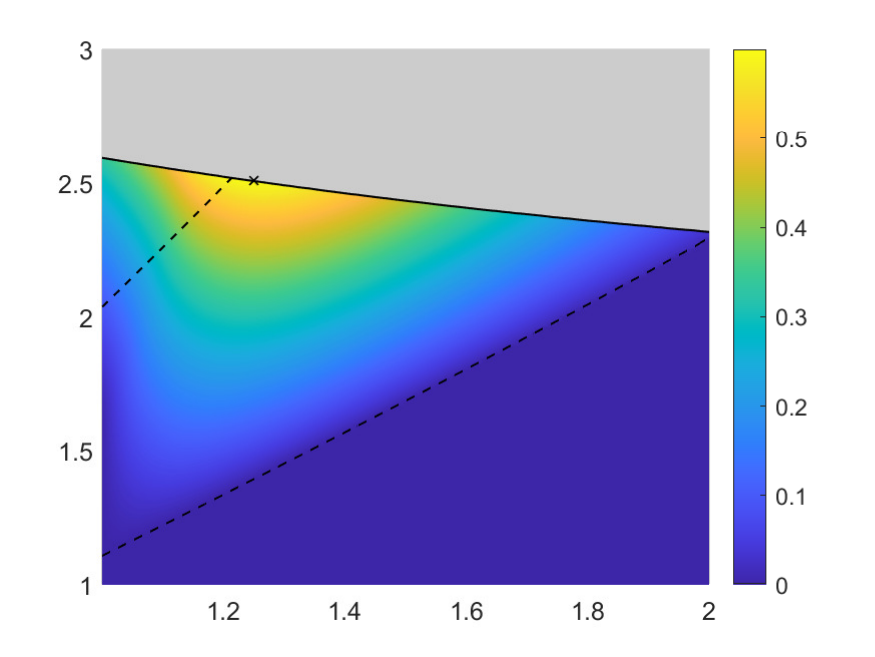}
\put(12,44){\color{white}\scriptsize\rotatebox{44}{snap-fit}}
\put(38,28){\color{white}\footnotesize\rotatebox{28}{stick-fit}}
\put(63,12){\color{white}\footnotesize{eject-fit}}
\put(66,64){\footnotesize{no-fit}}
\put(0,72){\small{(a)}}
\put(0,37.5){\footnotesize{$\phi$}}
\put(46,1){\footnotesize{$\alpha$}}
\end{overpic}
\end{subfigure}
\begin{subfigure}{0.9\columnwidth}
\hspace{0.5mm}
\begin{overpic}[width=1.1\linewidth,tics=10]{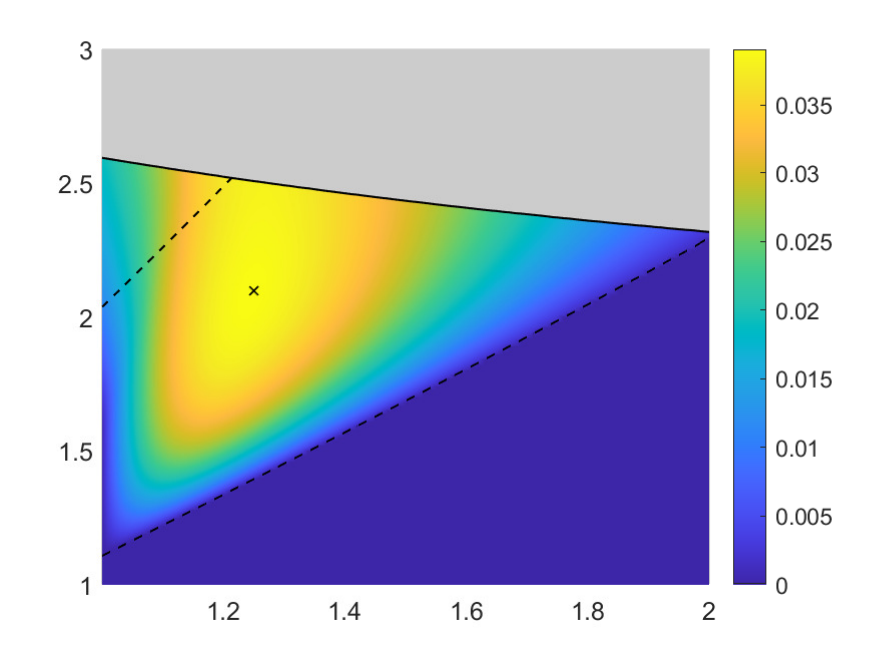}
\put(12,44){\color{white}\scriptsize\rotatebox{44}{snap-fit}}
\put(38,28){\color{white}\footnotesize\rotatebox{28}{stick-fit}}
\put(63,12){\color{white}\footnotesize{eject-fit}}
\put(66,64){\footnotesize{no-fit}}
\put(0,72){\small{(b)}}
\put(46,1){\footnotesize{$\alpha$}}
\end{overpic}
\end{subfigure}
\vspace{2mm}
\caption{Performance metrics of a grip-fit as the parameters $\alpha$ and $\phi$ vary. Colour maps show (a) the maximum lifting capacity $F_d^\mathrm{max}$, and (b) the maximum lifting capacity with fixed shell mass, $F_d^\mathrm{max}/\phi^3$, for $\mu=0.5$. The transitions between different fit regimes are marked by the dashed lines, with the no-fit boundary indicated by a solid line. The maximum in each plot is displayed by a cross ($\times$) at $(\alpha,\phi)\approx(1.25,2.51)$ and $(1.25,2.1)$, respectively. 
}\label{fig:FD_colourplots}
\end{figure}

\subsection{Optimizing with fixed shell mass}

Not only can an almost complete shell be impractical during assembly, but it will also require more material to be used in its manufacture --- a factor to consider in cost-effective manufacturing. Given that the typical force scale is proportional to the cube of the shell thickness, $h$, this leads to an interesting trade-off: at fixed mass, the product $h\phi$ is constant, so we expect that the dimensional force scale will be proportional to $\hat{F}_d\propto h^3 F_d(\phi)\propto F_d(\phi)/\phi^3$.
We therefore plot the rescaled lifting force $F_d^\mathrm{max}/\phi^3$ in fig.~\ref{fig:FD_colourplots}(b). This shows that a larger region seems to  optimize this  lifting force at fixed shell mass. Moreover, the global maximum now occurs away from the boundary with no-fit and, for $\mu \gtrsim 0.3$, occurs \emph{within} the stick-fit region. We also find that, while the value of $\alpha$ remains consistent, the value of $\phi$ at the global maximum decreases as $\mu$ increases. This shifts the previous result into something that is perhaps more anticipated; the suggestion that there is an advantage of taking a value of $\phi < \phi_{\text{no-fit}}$ that still maintains a high lifting capacity.

\subsection{The trade-off between assembly force and lifting capacity}

Whilst the lifting capacity may be the relevant priority in some scenarios, in other examples it is also a consideration that the assembly force not be too high, as might be the case for recreational purposes such as LEGO and Playmobil. To consider the trade-off between lifting and assembly, we first consider the assembly force, denoting the maximum  force during assembly by $F_a^\mathrm{max}=\max(F_a)$ (fig.~\ref{fig:FD_FA_colourplots}(a)). One can see the assembly force is largest when $\alpha$ is largest but $\phi$ is small. This is believed to be due to the eject-fit configurations, with large $\alpha$ requiring a lot of force to push the shell to the cylinder while overcoming the repelling force. Nothing illuminating is seen with regards to minimizing the assembly force alone -- this is simply equivalent to taking the smallest value of $\alpha$. 

Figure \ref{fig:FD_FA_colourplots}(b) shows the behaviour of the ratio of forces, $F_d^\mathrm{max}/F_a^\mathrm{max}$, as a measure of lifting capacity at the cost of assembly, which we refer to as the `locking gain'. Here, we see a dramatic shift in the geometric parameters for which performance is optimized. As anticipated from the typical characteristics of a snap-fit connection, for all values of $\mu$, the optimal region has $F_d^\mathrm{max}/F_a^\mathrm{max}>1$. This quantifies what we know  from everyday experience of snap-fits: that a well-designed snap-fit connection is easier to assemble than disassemble. To highlight the maximum of this ratio as $\phi$ varies below the no-fit boundary, fig.~\ref{fig:FD_FA_colourplots} shows the corresponding value of $\alpha$ for which the global maximum of $F_d^\mathrm{max}/F_a^\mathrm{max}$ is obtained. 
The behaviour of the optimum region and optimum curve remains similar for other values of $\mu$, with small shifts in absolute values. Whilst this optimum curve seems to follow alongside the no-fit boundary for larger values of $\phi$, one can confirm that the optima do not lie on the boundary.

\begin{figure}[]
\begin{subfigure}{0.9\columnwidth}
\hspace{-2.5mm}
\begin{overpic}[width=1.1\linewidth,tics=10]{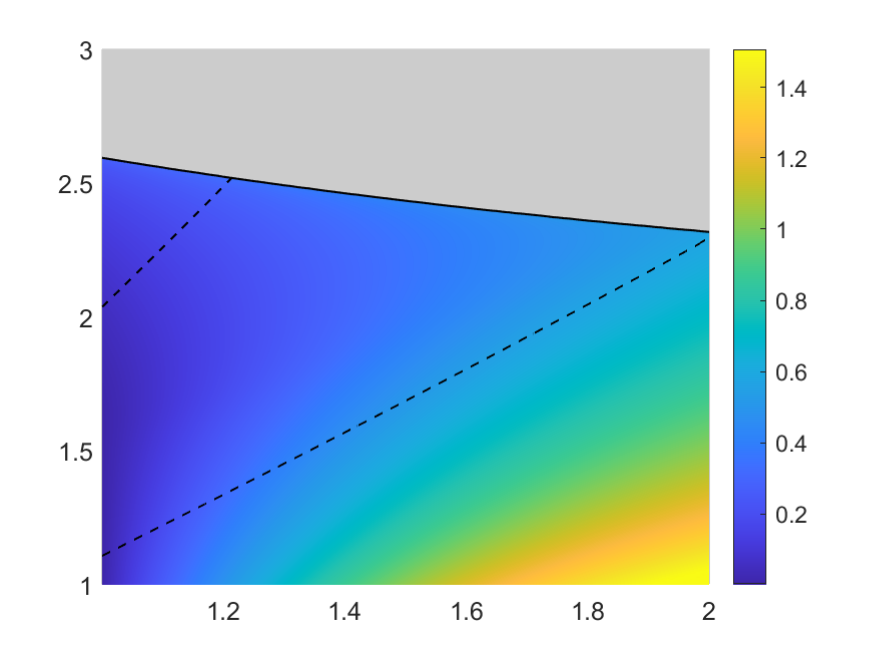}
\put(12,44){\color{white}\scriptsize\rotatebox{44}{snap-fit}}
\put(38,28){\color{white}\footnotesize\rotatebox{28}{stick-fit}}
\put(24,12){\color{white}\footnotesize{eject-fit}}
\put(66,64){\footnotesize{no-fit}}
\put(2,37.5){\footnotesize{$\phi$}}
\put(46,2){\footnotesize{$\alpha$}}
\put(0,72){\small{(a)}}
\end{overpic}
\end{subfigure}
\begin{subfigure}{0.9\columnwidth}
\hspace{-2mm}
\begin{overpic}[width=1.1\textwidth,tics=10]{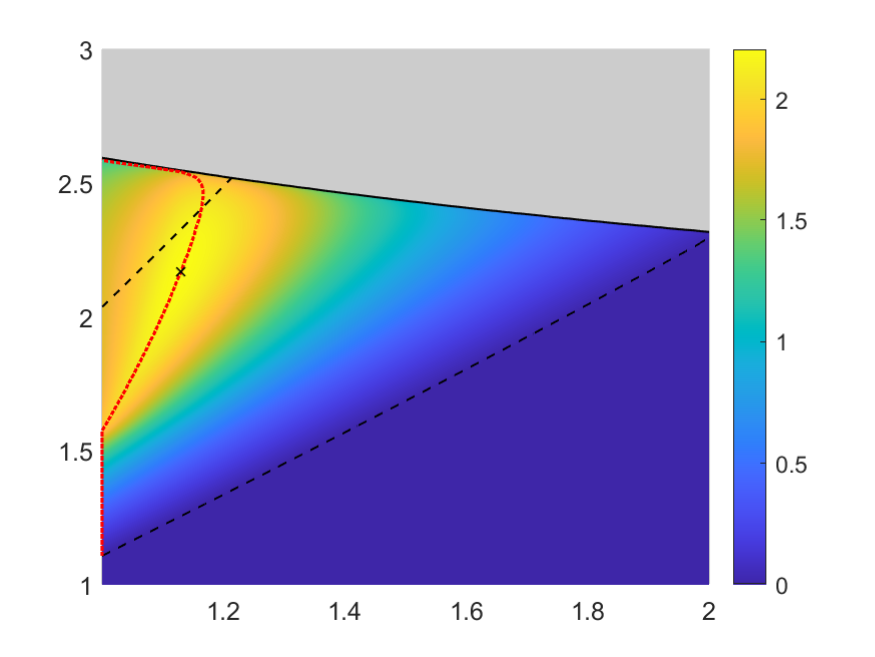}
\put(12,44){\scriptsize\rotatebox{44}{snap-fit}}
\put(38,28){\color{white}\footnotesize\rotatebox{28}{stick-fit}}
\put(63,12){\color{white}\footnotesize{eject-fit}}
\put(66,64){\footnotesize{no-fit}}
\put(46,2){\footnotesize{$\alpha$}}
\put(0,72){\small{(b)}}
\end{overpic}
\end{subfigure}
\vspace{2mm}
\caption{(a) Colour map of the maximum assembly force $F_a^\mathrm{max}(\alpha,\phi)$ with $\mu=0.5$. (b) Colour map of the maximum lifting-to-assembly force $F_d^\mathrm{max} / F_a^\mathrm{max}$ for opening angles $\phi$ against ratio of radii $\alpha$, for $\mu=0.5$. The optimum $\alpha$ for a given $\phi$ is shown by the red dotted curve, with the global maximum displayed by a cross ($\times$) at $(\alpha,\phi)\approx(1.13,2.17)$. The transitions between different fit regimes are labelled and marked by dashed lines, while the no-fit boundary is denoted by a solid line. }\label{fig:FD_FA_colourplots}
\end{figure}
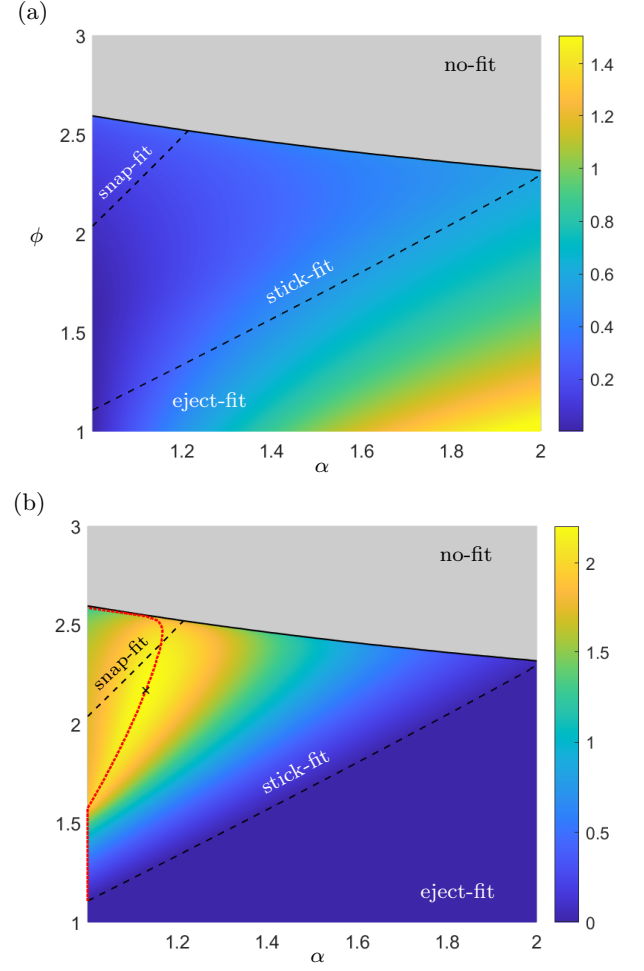

\subsection{Lifting fragile objects}

In addition to raw lifting capability and a trade-off of disassembly versus ease of assembly, another performance consideration may be control of the clamping force exerted by the gripper onto the object. For example, we may be seeking to grip a cylinder that is rigid up to a given applied normal force, but beyond which the material may deform or be damaged by the force of the gripper, e.g.~fitting an egg into an `egguin'\footnote{A cylindrical gripper to hold  multiple eggs while cooking, see \texttt{https://peleg-design.com/products/egguins} .}. In addition, while eggs are relatively strong when subject to evenly distributed external forces, they are vulnerable to `pinching' or `poking' forces. Consequently, another parameter to monitor when choosing an optimum configuration is the normal reaction force $F_N$. 
We denote the normal reaction force for assembly and disassembly as $F_{dN}$ and $F_{aN}$,  respectively. We note that, in all cases considered, the maximum normal reaction force arises during disassembly, although we do not have that $F_{dN}>F_{aN}$ in general. The maximum clamping force, $F_N^\mathrm{max}$ is given by $$F_N^\mathrm{max}=\max(F_{dN})=\max\left(\frac{F_d}{2} \frac{1}{\cos\psi + \mu \sin\psi} \right).$$ 

The value of $F_N^\mathrm{max}$ appears to follow the same pattern for each value of $\mu$: the greatest increase in the clamping force occurs with increasing $\alpha$, while only a relatively slight increase comes from increasing $\phi$ (fig.~\ref{fig:clamping_force}(a)). This indicates that minimizing $\alpha$ will minimize the clamping force, hence reinforcing the idea that the optimum configuration occurs when the shell and cylinder have similar radii.

\begin{figure}[]
\begin{subfigure}{0.9\columnwidth}
\hspace{-3.5mm}
\begin{overpic}[width=1.1\linewidth,tics=10]{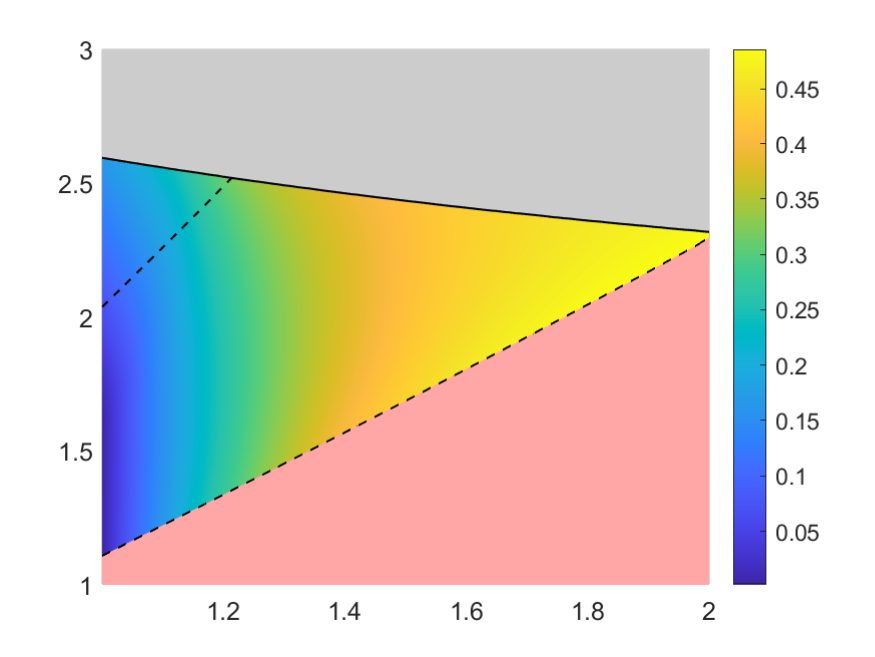}
\put(11.75,44.25){\color{white}\scriptsize\rotatebox{44}{snap-fit}}
\put(35,27.5){\color{black}\footnotesize\rotatebox{28}{stick-fit}}
\put(63,12){\color{black}\footnotesize{eject-fit}}
\put(66,64){\footnotesize{no-fit}}
\put(0,37.5){\footnotesize{$\phi$}}
\put(46,2){\footnotesize{$\alpha$}}
\put(0,72){\small{(a)}}
\end{overpic}
\end{subfigure}
\begin{subfigure}{0.9\columnwidth}
\hspace{-2mm}
\begin{overpic}[width=1.1\linewidth,tics=10]{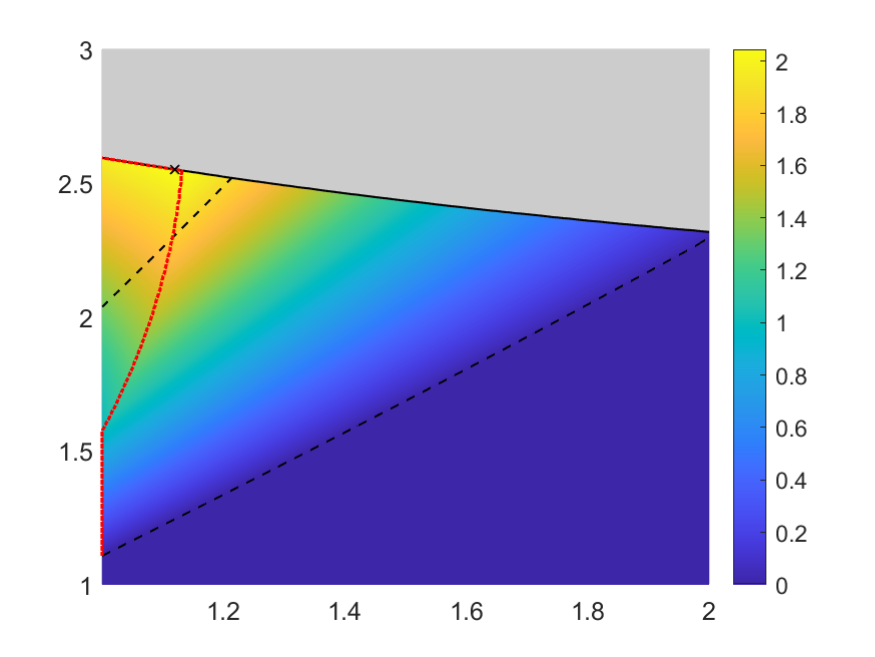}
\put(11.75,44.25){\scriptsize\rotatebox{45}{snap-fit}}
\put(35,27.25){\color{white}\footnotesize\rotatebox{30}{stick-fit}}
\put(63,12){\color{white}\footnotesize{eject-fit}}
\put(66,64){\footnotesize{no-fit}}
\put(46,2){\footnotesize{$\alpha$}}
\put(0,72){\small{(b)}}
\end{overpic}
\end{subfigure}
\vspace{2mm}
\caption{(a) Colour plot of the maximum clamping force $F_N^\mathrm{max}$ for opening angles $\phi$ against ratio of radii $\alpha$, for $\mu=0.5$. (b) Colour plot of the maximum lifting-to-clamping force $F_d^\mathrm{max}/F_N^\mathrm{max}$ for opening angles $\phi$ against ratio of radii $\alpha$, for $\mu=0.5$. The optimum $\alpha$ for a given $\phi$ is shown by the red dotted curve, with the overall maximum displayed by a cross ($\times$) at $(\alpha,\phi)\approx(1.12,2.55)$.}\label{fig:clamping_force}
\end{figure}
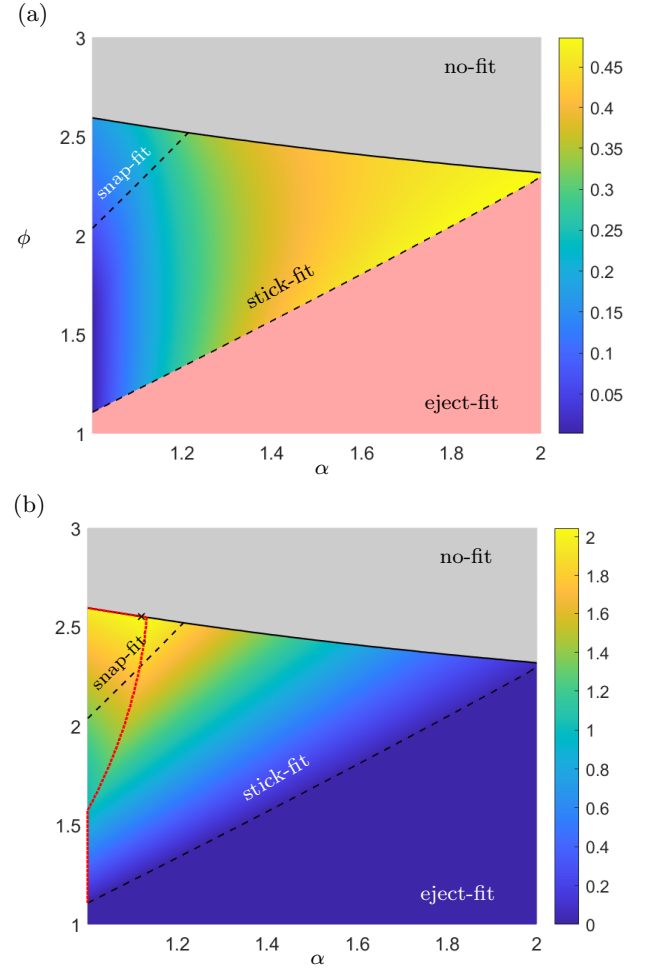

Combining the effects of both the lifting force and clamping force, by considering $F_d^\mathrm{max} / F_N^\mathrm{max}$, we find a similar optimal region to that found when seeking to maximize the lifting force alone (fig.~\ref{fig:clamping_force}(b)). The optimum curve behaves similarly to that for $F_d^\mathrm{max}/F_a^\mathrm{max}$ except this time the optimum follows the no-fit boundary above a critical value of $\phi$. Whilst we see the overall maximum is back near the no-fit boundary, this optimum region of larger force now branches down into smaller values of $\phi$ and reaches $\alpha \gtrsim1$, suggesting once again that it is beneficial to choose smaller $\alpha$, but that there is still flexibility in the choice of $\phi$. Thus, our analysis indicates that a snap-fit gripper whose radius is comparable to the object being lifted is likely to cause the least amount of damage to the object.

\section{Conclusions}

In this article, we have investigated the deformation and behaviour of an elastic shell used to lift a cylindrical load. We have shown that a new regime --- the stick-fit --- is possible and that this can outperform the snap-fit considered previously \cite{yoshida2020snap}.

We have characterized the  geometrical properties of the gripper under which each of four different behaviours are observed: \emph{snap-fit}, where the shell is drawn onto the cylinder; \emph{stick-fit}, where the shell is retained on the cylinder without snapping; \emph{eject-fit}, where the shell is repelled from the cylinder; and \emph{no-fit}, where the shell does not slide around the cylinder at all. We produced a regime diagram (fig.~\ref{fig:snap_regime_diagram}) that captures the effects of three control parameters: the opening angle of the shell, $\phi$, the ratio of radii of the cylinder to the undeformed shell, $\alpha$, and the coefficient of friction, $\mu$. We found that, while the range of geometrical parameters that produce a no-fit increases with $\mu$, the collective grip-fit regime expands with $\mu$. This suggests that the range of shells capable of lifting increases with friction: a shell producing high friction would possess a nonzero lifting capacity for a larger range of cylindrical radii. 

Our main interest lies in the snap-fit and stick-fit regimes (collectively referred to as `grip-fit') in which the shell has a nonzero lifting capacity. 
We therefore investigated the lifting capacity of the shell in the grip-fit regime. We discussed and  analysed various criteria by which the lifting might be considered optimal: raw lifting ability, best lifting ability at fixed amount of shell material, maximum `locking gain', and lifting without damaging the load.  In each case, we found that there exists a broad range of geometrical parameters of the shell (its opening angle and radius) that are close to optimal. In each case, global maxima exist with a shell radius that is only $10$--$20\%$ smaller than that of the load. (This is consistent with designs commonly used for pipe clamps, in which the inner radius of the clamp closely matches that of the pipe.) Somewhat surprisingly, we found that, with sufficiently large friction coefficients, these global maxima may exist within the stick-fit, rather than snap-fit, regimes.

Although all metrics suggest choosing comparable shell and cylinder radii, there is no clear advantage to using identical radii rather than a slightly smaller shell. The discrepancy between this and the design of pipe clamps, which favour identical radii, may arise because we are modelling a thin shell with point contact rather than a gripper with finite thickness and extended regions of contact. Further, in reality, snap-fit components are commonly manufactured with plastic, therefore the best choice of shell when manufacturing would require minimal stress in the assembled state --- periods of excessive stress can cause the plastic to age.

Inspired by the consideration of a cylindrical snap-fit as a simple example of a two-claw soft-gripper, our work is centred around the lifting of a cylindrical object. Whilst the cylindrical snap-fit is perhaps not as versatile as a two-claw gripper (it is most effective with cylindrical objects) it nevertheless has a role because of its speed, security, reversibility and the fact that it operates passively. To further this work and increase versatility, one could consider relaxing this assumption of a cylindrical object  and consider how this lifting capacity changes with an object of a different shape or an imperfect cylinder.

\begin{acknowledgments}
For the purpose of open access, the authors have applied a CC BY public copyright licence to any author-accepted manuscript (AAM) arising from this submission.
\end{acknowledgments}


\vskip2pc



%

\end{document}